\documentclass[12pt,3p,authoryear]{elsarticle}
\usepackage[T1]{fontenc}
\usepackage{amsmath,amssymb,amsfonts}
\usepackage{natbib}
\usepackage{graphicx}
\usepackage[hyphens]{url}
\usepackage[table]{xcolor}
\usepackage{epstopdf}
\usepackage{booktabs}
\usepackage{rotating}
\usepackage{times}
\usepackage{bm}
\usepackage{xspace}
\usepackage{color}
\usepackage{subdepth}
\usepackage{float}
\usepackage{setspace}
\usepackage{needspace}   
\usepackage[labelfont=bf,position=top]{caption}
\usepackage{geometry}
\usepackage[%
pdftex,%
pdfauthor={Me}%
pdfstartview={XYZ null null 1},%
pdffitwindow=true,%
pdfcreator={pdflatex}
letterpaper,%
colorlinks,%
breaklinks=true,%
backref=true,%
plainpages=false,%
bookmarks=false,%
pdfpagelabels,
pdfpagemode=None%
]{hyperref}

\usepackage[font={small,singlespacing},labelfont={bf,small},skip=0.2cm]{caption}
\floatstyle{plaintop}
\restylefloat{table}
\restylefloat{figure}

\RequirePackage{ae,fancyvrb}

\journal{SSRN}

\usepackage{lineno}

\usepackage{adjustbox}
\usepackage{longtable}
\usepackage{tabularx,xltabular}

\newcommand{\eg}{\emph{e.g.}\xspace}

\newcommand{\insertfloat}[1]{%
\begin{center}
[Insert~#1 about here.]%
\end{center}%
}

\begin{document}
\begin{frontmatter}
\title{The Role of Twitter in Cryptocurrency Pump-and-Dumps\tnoteref{label1}}
\tnotetext[label1]{We acknowledge the \emph{Centre d’intelligence en surveillance des marchés financiers} for its financial support.}
\author[hec]{David Ardia}
\ead{david.ardia@hec.ca}
\author[sher]{Keven Bluteau\corref{cor1}}
\ead{keven.bluteau@usherbrooke.ca}
\cortext[cor1]{Corresponding author. Université de Sherbrooke, École de Gestion, 2500 Boulevard de l’Université, Sherbrooke J1K 2R1. Phone: +1 514 340 6103.}
\address[hec]{GERAD \& Department of Decision Sciences, HEC Montréal, Canada}
\address[sher]{Department of Finance, Université de Sherbrooke, Canada}

\begin{abstract}
We examine the influence of Twitter promotion on cryptocurrency pump-and-dump events. By analyzing abnormal returns, trading volume, and tweet activity, we uncover that Twitter effectively garners attention for pump-and-dump schemes, leading to notable effects on abnormal returns before the event. Our results indicate that investors relying on Twitter information exhibit delayed selling behavior during the post-dump phase, resulting in significant losses compared to other participants. These findings shed light on the pivotal role of Twitter promotion in cryptocurrency manipulation, offering valuable insights into participant behavior and market dynamics.
\end{abstract}
\end{frontmatter}

\clearpage
\setcounter{page}{1}
\doublespacing

\section{Introduction}


\noindent
The rapid growth of the cryptocurrency market has attracted significant attention from investors, regulators, and researchers. As the market evolves, new forms of market manipulation emerge, posing challenges to market integrity and investor protection. One such manipulation strategy that has gained prominence is the ``pump-and-dump'' scheme, where a targeted cryptocurrency's price is artificially inflated through coordinated buying, followed by a sudden sell-off to unsuspecting investors. In recent years, the rise of social media platforms, particularly Twitter, has provided an avenue for promoting and coordinating these pump-and-dump events \citep[see, \eg,][]{li2021cryptocurrency,NizzoliEtAl2020,MirtaheriEtAl2021}. Due to the relatively young age of the cryptocurrency market, Twitter is also one of the primary sources of (legitimate) information for cryptocurrency investors \citep[see, \eg,][]{KraaijeveldDeSmedt2020}. 

This research paper investigates the role of Twitter promotion in pump-and-dump events within the cryptocurrency market. By examining the interplay between abnormal returns, trading volume, and tweets, we aim to shed light on the dynamics of these events and their relationship with social media activity. Specifically, we seek to determine the impact of Twitter promotion on attracting participants to pump-and-dump events and its influence on participant behavior before, during, and after the events. 

Most of the efforts of the literature have been regarding the identification of pump-and-dump events using statistical and machine-learning methodologies. Most approaches use time-series anomaly detection \citep[see, \eg,][]{KampsKleinberg2018,MansourifarEtAl2020}, or classification models based on market features \citep[see, \eg,][]{VictorHagemann2019,XuLivshits2019}, and social media data \citep[see, \eg,][]{NghiemEtAl2021}. Recently \citet{li2021cryptocurrency} analyze the behavior of price, volume and volatility around cryptocurrency pump-and-dump events while \citet{DhawanPutnins2023} propose a theorical framework on why people participate in cryptocurrencies pump-and-dump scheme. However, to our knowledge, no study has investigated the role of Twitter promotion on participant behavior around pump-and-dump events. We employ a comprehensive methodology that combines graphical analysis and regression modeling to accomplish our research objectives. Our analysis begins with visually examining the cumulative abnormal return, volume, and tweets, providing a visual depiction of their evolution leading up to, during, and following pump-and-dump events. This graphical analysis serves as a foundation for our subsequent regression analysis, which proceeds in three steps. 

First, we investigate the contemporaneous relationship between the abnormal number of tweets before the event and the abnormal return before the event. By focusing on the last 30 minutes of the pre-event window, we aim to determine if increased Twitter activity corresponds to higher abnormal returns. Consistent with the notion that promotion on Twitter engenders heightened attention towards the targeted cryptocurrency, subsequently generating increased buying pressure, we observe a positive and statistically significant relationship between the abnormal number of tweets and abnormal returns. 

Second, we delve deeper into the dynamics of pump-and-dump events by examining the relationship between the cumulative abnormal number of tweets before the event and abnormal returns during the dump phase. We explore two scenarios: one where investors informed via Twitter are aware of the impending dump and adjust their behavior accordingly, and another where they are unaware and their behavior remains unaffected. Our findings reveal a significant negative relationship between the cumulative abnormal return at and preceding the event and its equivalent during the dump window. This indicates that a substantial pump event corresponds to a sizable dump event. However, we do not observe a significant relationship between the abnormal number of tweets before the event and the cumulative abnormal return during the dump window. This outcome suggests that investors informed via Twitter are not selling their investments during the dump period.

Finally, we explore the relationship between the cumulative abnormal number of tweets before the event and abnormal returns during the post-dump phase. We observe a predominantly negative and statistically significant relationship between the post-dump abnormal return and the cumulative abnormal number of tweets before the event. This indicates that the price increase observed before the event, which is associated with abnormal tweets, only reverses at least 30 minutes after the dump phase. This finding suggests that investors who receive information from Twitter regarding a pump-and-dump event tend to be delayed in selling their cryptocurrencies compared to investors who receive information through other channels related to the event's announcement. Moreover, we find that the reversal is substantially larger than the contemporaneous effect, implying that investors informed via Twitter who are late in selling their cryptocurrencies during a pump-and-dump event tend to experience, on average, a loss on their investment.

The rest of the paper is as follows. Section~\ref{sec:anatomy} presents 
an overview of pump-and-dump groups and events. Section~\ref{sec:data} describes the various datasets used in the study. We present the methodology and the empirical results in Section~\ref{sec:empirical}. Section~\ref{sec:conclusion} concludes.

\section{Pump-and-Dump Groups and Events}
\label{sec:anatomy}

\noindent
In this section, we introduce pump-and-dump groups and events in cryptocurrencies. 

\subsection{Telegram}

\noindent
The utilization of Telegram's platform (\url{https://telegram.org/}) by organizers engaged in cryptocurrency pump-and-dump practices serves as a means to coordinate their activities and cultivate a substantial community \citep[see, \eg,][]{LaMorgiaEtAl2020,HamrickEtAl2021,MirtaheriEtAl2021}. Telegram was launched in 2013 as an instant messaging service and has since garnered a user base exceeding 700 million worldwide as of June 2022.\footnote{See \url{https://telegram.org/blog/700-million-and-premium}.} This platform, available free of charge, operates on a cloud infrastructure and provides robust data protection measures, ensuring the security of user information. Moreover, Telegram is accessible across various devices, making it highly convenient for users. It facilitates communication through individual messaging, group chats accommodating up to 200,000 users, and channels with unlimited subscribers. These features make Telegram a preferred choice for organizing market manipulation activities such as pump-and-dumps \citep[see, \eg,][]{li2021cryptocurrency}.\footnote{It is pertinent to acknowledge that Discord, another similar platform, is also widely used for cryptocurrency pump-and-dump schemes, albeit predominantly among the gaming community. Due to its greater user base, our study focuses solely on Telegram.}

While most groups on Telegram can be accessed without restrictions, granting users access to comprehensive message histories along with metadata such as timestamps and the number of views for each message, certain groups and channels operate on an invite-only basis. Furthermore, Telegram allows for the deletion of entire groups and individual messages, posing challenges in reconstructing the complete historical ecosystem pertaining to pump-and-dump operations within the cryptocurrency context.

\subsection{Pump-and-Dump Groups}

\noindent
The formation of cryptocurrency pump-and-dump groups necessitates many participants to achieve significant price movements. However, assembling such a community is a complex and time-consuming process. In the early stages, these groups typically share signals from more established counterparts. Group members then promote their group on platforms like Reddit or within existing groups. Once a certain threshold of members is reached, the group can begin organizing events while striving for continued growth.

Within a pump-and-dump group, a hierarchical structure emerges. Administrators hold privileged information and serve as the primary communicators, exclusively posting messages. Participants seek any pertinent updates, while VIP members, who often pay a subscription fee, receive information slightly earlier through an invite-only channel.

Administrators regularly disseminate promotional messages, share past event returns using price curves, offer advice to newcomers, and focus on a specific exchange to maximize potential gains. To initiate a pump-and-dump, administrators share an announcement specifying the date, time, and targeted exchange. They continuously remind participants to engage, promote the group, and encourage VIP membership. Eventually, they distribute the signal, typically mentioning only the name of the cryptocurrency. Administrators benefit most from the pump-and-dump by possessing advanced information, allowing them to enter the market ahead of VIP and regular users, maximizing profits.

Group members await the signal, prepared to trade their tokens on the designated platform. Prompt execution is crucial. However, many participants need more time to enter the market, purchasing inflated cryptocurrencies and often struggling to sell them promptly (with the event) and thus face losses. Ultimately, group members bear the brunt of losses associated with participating in pump-and-dump activities.

VIP members pay a subscription fee to gain early access to information, sometimes receiving it more than 24 hours before regular members, which affords them a significant advantage. VIPs also gain access to exclusive events inaccessible to others. In addition, in certain groups, non-subscribing participants can attain different membership levels by inviting a specified number of people.

The organizers of pump-and-dump activities select exchanges for their operations. Popular choices include Binance, Bittrex, Coinbase, Hotbit, and KuCoin, with Binance being the most frequently utilized. Despite the negative consequences of pump-and-dump schemes, exchanges have little incentive to ban them. Such events generate high transaction volumes, leading to increased costs for exchanges. Additionally, exchanges often hold cryptocurrencies that can yield profits post-pump-and-dump while attracting more users to their platforms. Furthermore, exchanges possess important user information, enabling them to monitor participants or engage in insider trading during future pump-and-dumps. For instance, YoBit openly admitted to engaging in token pumping and dumping, earning the moniker of the ``king of pump-and-dumps'' in October 2018.\footnote{Refer to \url{https://fullycrypto.com/like-it-or-not-yobit-is-still-the-pump-and-dump-king} for details.}

\subsection{Pump-and-Dump Events}

\noindent
The initial step in a pump-and-dump scheme entails the event's announcement, typically occurring a few days or even just a few hours beforehand. A longer interval between the announcement and the signal allows group members to prepare their capital for the opportune moment. The announcement includes the specific date and time of the event, along with the targeted exchange. Furthermore, it is accompanied by motivational and promotional messages designed to attract a more exclusive circle of users.

\begin{figure}[H]
\caption{\textbf{Example of a Pump-and-Dump Event}\\
This figure displays a series of messages from the ``Big\_Pumps\_Signals'' channel on Telegram, showcasing the announcement for a pump-and-dump event involving the cryptocurrency ``DREP'' on July 25, 2021, at 5 PM GMT on the Binance exchange.}
\label{f:exP&D}
\centering	
\includegraphics[scale=0.75]{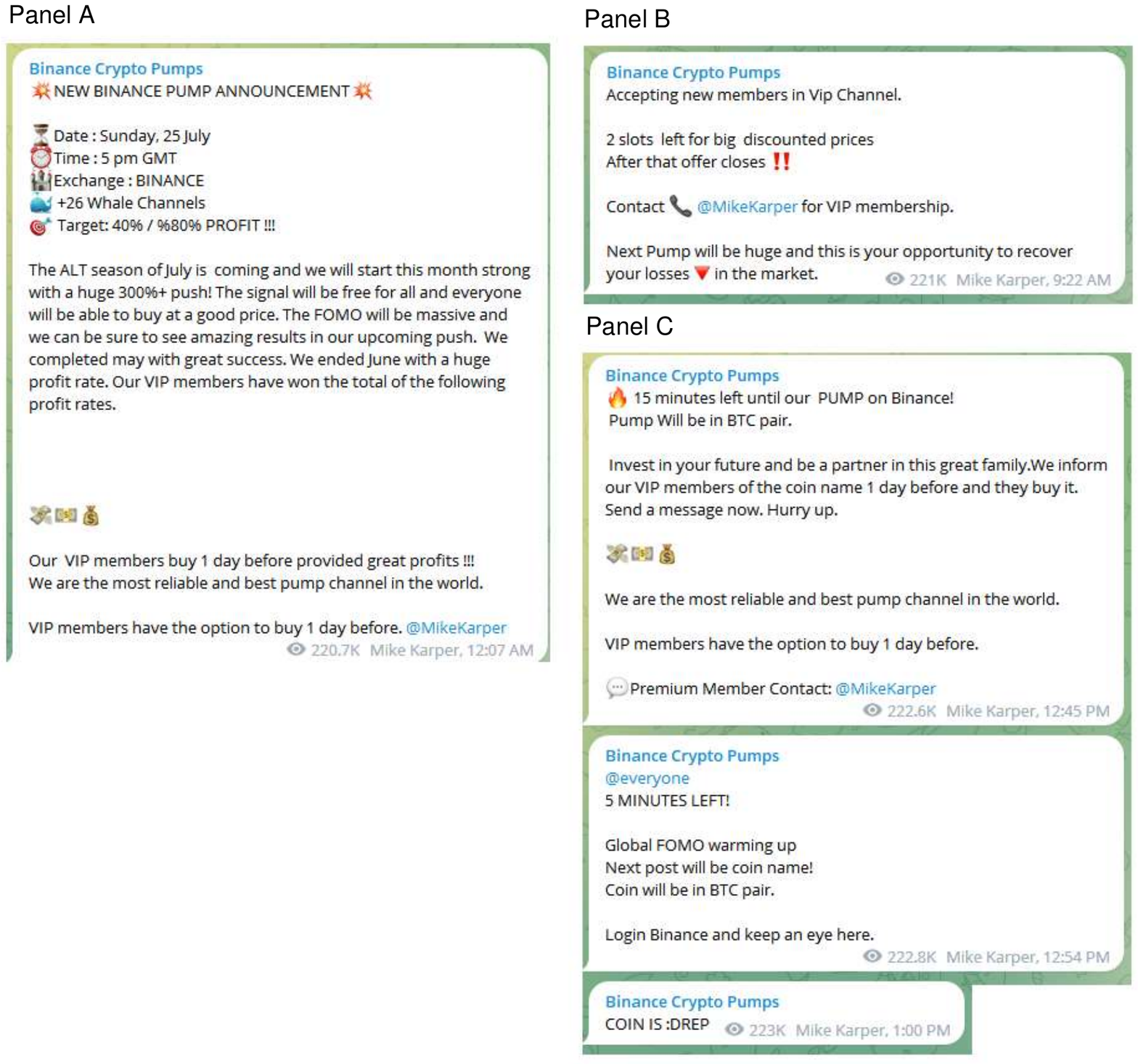}
\end{figure}

For example, Figure~\ref{f:exP&D} illustrates a series of messages from the ``Big\_Pumps\_Signals'' channel on Telegram, showcasing the announcement for a pump-and-dump event involving the cryptocurrency ``DREP'' on July 25, 2021, at 5 PM GMT on the Binance exchange. The first mention corresponds to the announcement of a new pump, issued several days or even hours before the event, as observed in this case, four days beforehand on July 21, 2021. The announcement specifies the event's date, time, and targeted exchange.

\insertfloat{Figure~\ref{f:exP&D}}

\begin{figure}[H]
\caption{\textbf{Example of VIP Access Advantages}\\ 
This figure highlights the benefits of VIP membership (in this instance, retrieved from the ``BestBinancePumps'' Telegram channel). This screenshot was taken directly from the VIP member link in the Telegram channel, which, at the time of this writing, was available at \url{https://shorturl.at/hkSUZ}. We note that this page changes over time.}
\label{f:VIPmemberExample}
\centering
\includegraphics[scale=0.7]{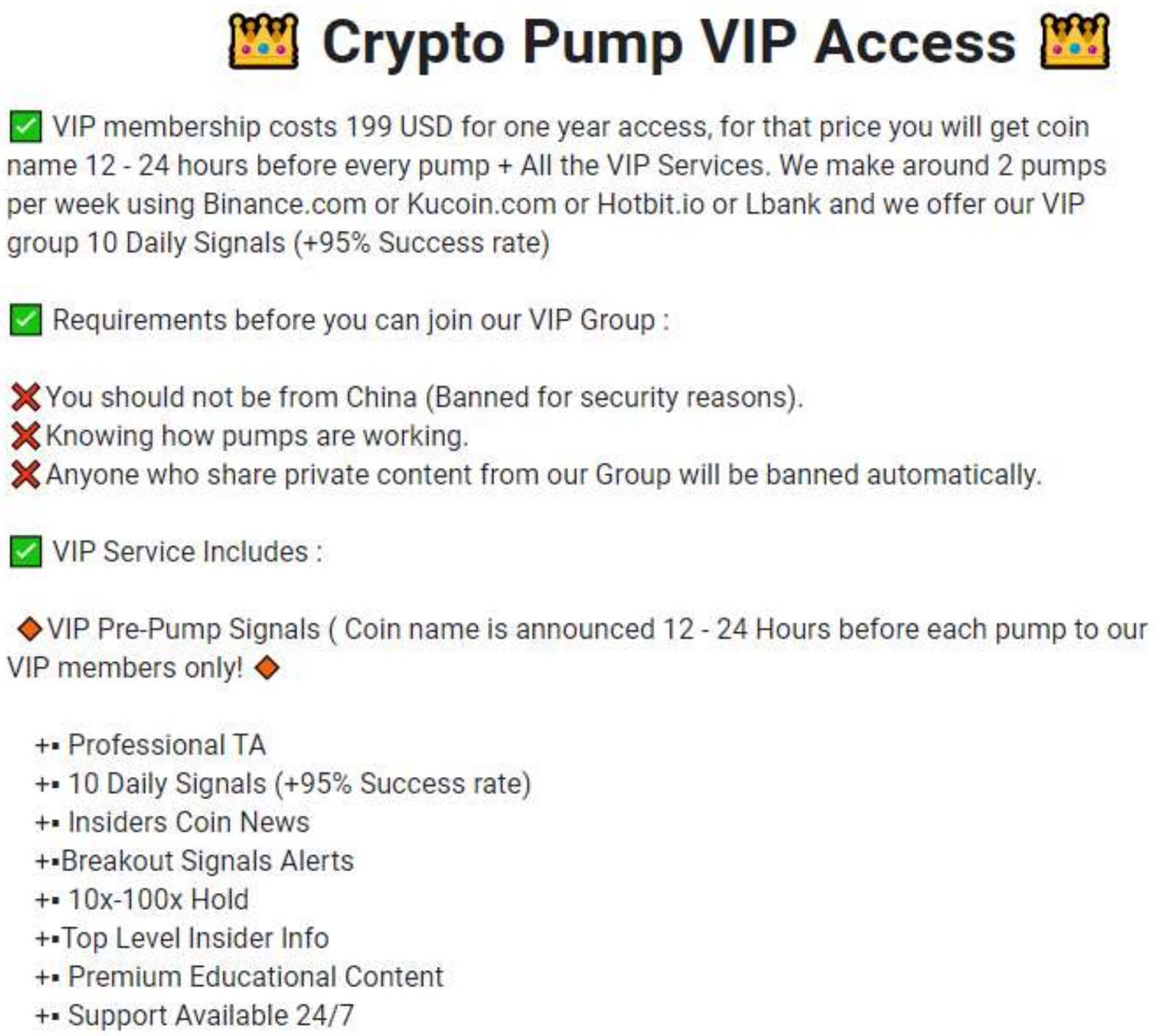}
\end{figure}

In addition to the announcement, promotional messages may be reiterated leading up to the event. As shown in Figure~\ref{f:VIPmemberExample} with an example that highlights the benefits of VIP membership (in this instance, for the ``BestBinancePumps'' channel).\footnote{At the time of writing, the complete VIP member form was accessible at: \url{https://shorturl.at/hkSUZ}} It reveals that VIP members gain access to the coin name 12-24 hours before the pump signal release.

\insertfloat{Figure~\ref{f:VIPmemberExample}}

\begin{figure}[H]
\caption{\textbf{One Event on Multiple Groups}\\ 
This figure is a screenshot of Telegram search functionality when looking for the ``DREP'' pump-and-dump event on July 25th, 2021, among the various pump-and-dump groups considered in this study.}
\label{f:PDmulitplegroups}
\centering
\includegraphics[scale=0.65]{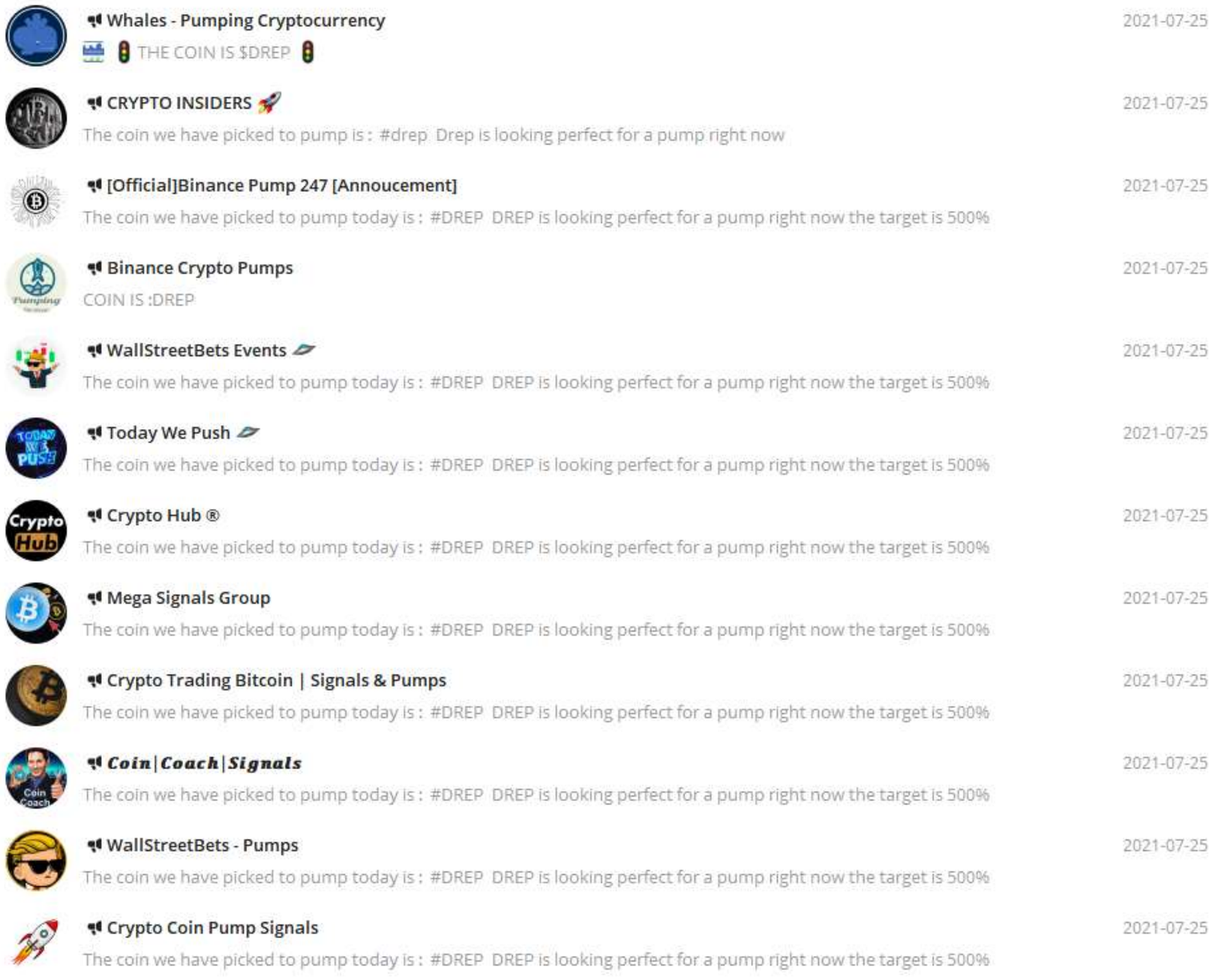}
\end{figure}

Finally, at the specified event time, the coin name is announced. A single event can be shared across multiple groups, as observed in the example above detected in at least 12 groups used in this study, as demonstrated in Figure~\ref{f:PDmulitplegroups}.

\insertfloat{Figure~\ref{f:PDmulitplegroups}}

\begin{figure}[H]
\caption{\textbf{Market Reaction Around a Successful Pump-and-Dump Event}\\
This figure reports the cumulative abnormal return and the cumulative abnormal volume of ``DREP'' around its July 25th, 2021, pump-and-dump events. Minute 0 is the exact announcement time reported on the ``Big\_Pumps\_Signals'' channel, at 5 pm GMT.}
\label{f:DERP_events}
\centering
\includegraphics[scale=0.35]{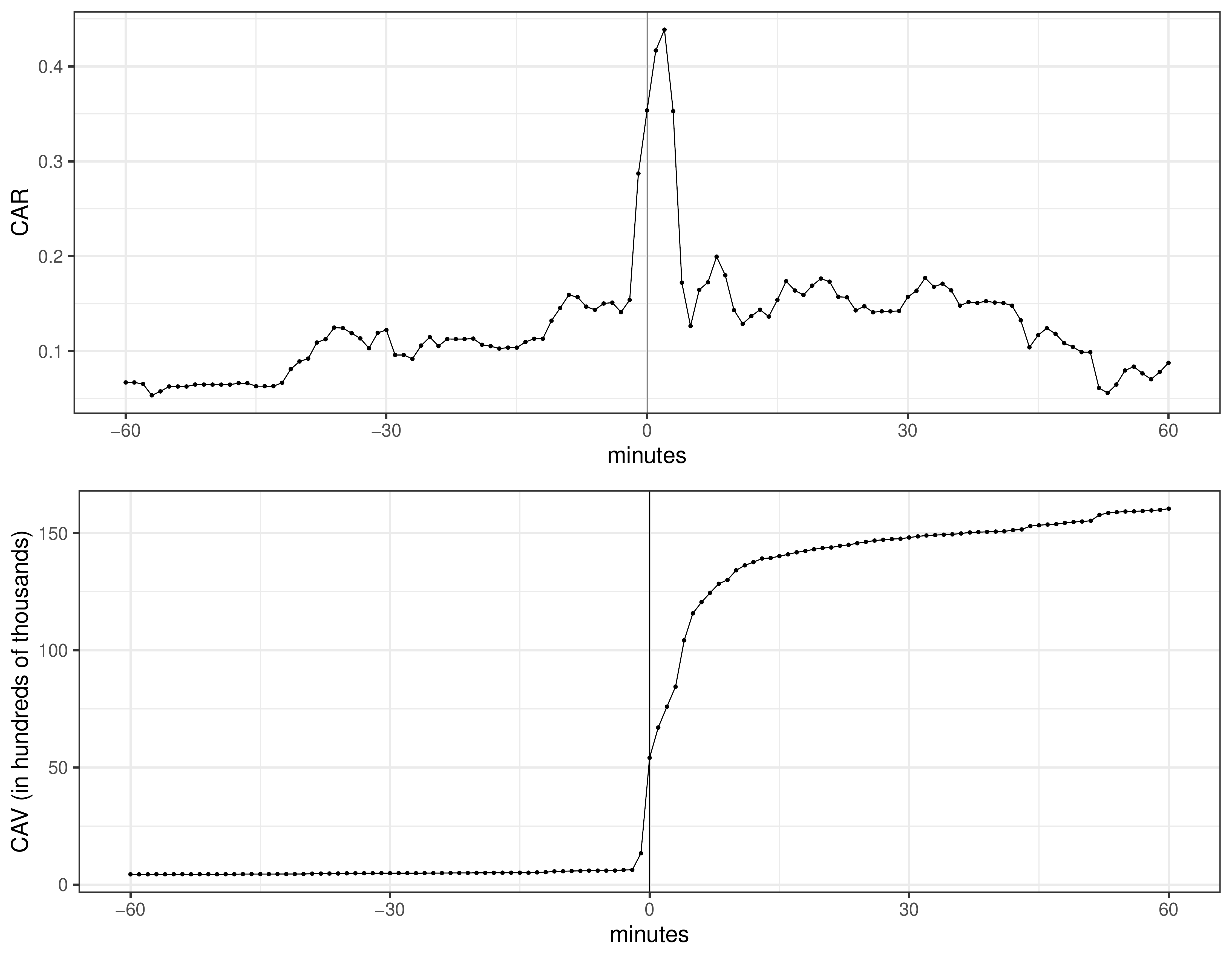}
\end{figure}

Figure~\ref{f:DERP_events} showcases the cumulative abnormal returns and cumulative abnormal volume (refer to Section~\ref{sec:methodology}) of the DREP events. The graph illustrates the remarkable success of this event, with cumulative abnormal returns reaching 35.36\% at the peak (at time 0). A substantial surge in volume accompanies the price reaction. Moreover, there were notable increases in returns before the event, indicating the involvement of VIP members, as evidenced by the preceding volume activity. However, these increases are minimal (or at least more gradual) compared to the surge observed during the event.

\insertfloat{Figure~\ref{f:DERP_events}}

\begin{figure}[H]
\caption{\textbf{Example of an Automated Pump Detection Service}\\
This figure showcases an example of the DREP event detected by the automated pump detection service provided in the channel ``cointrendz\_pumpdetector''.}
\hspace*{3.3cm}
\label{f:PumpDetectorDREP}
\centering
\includegraphics[scale=0.55]{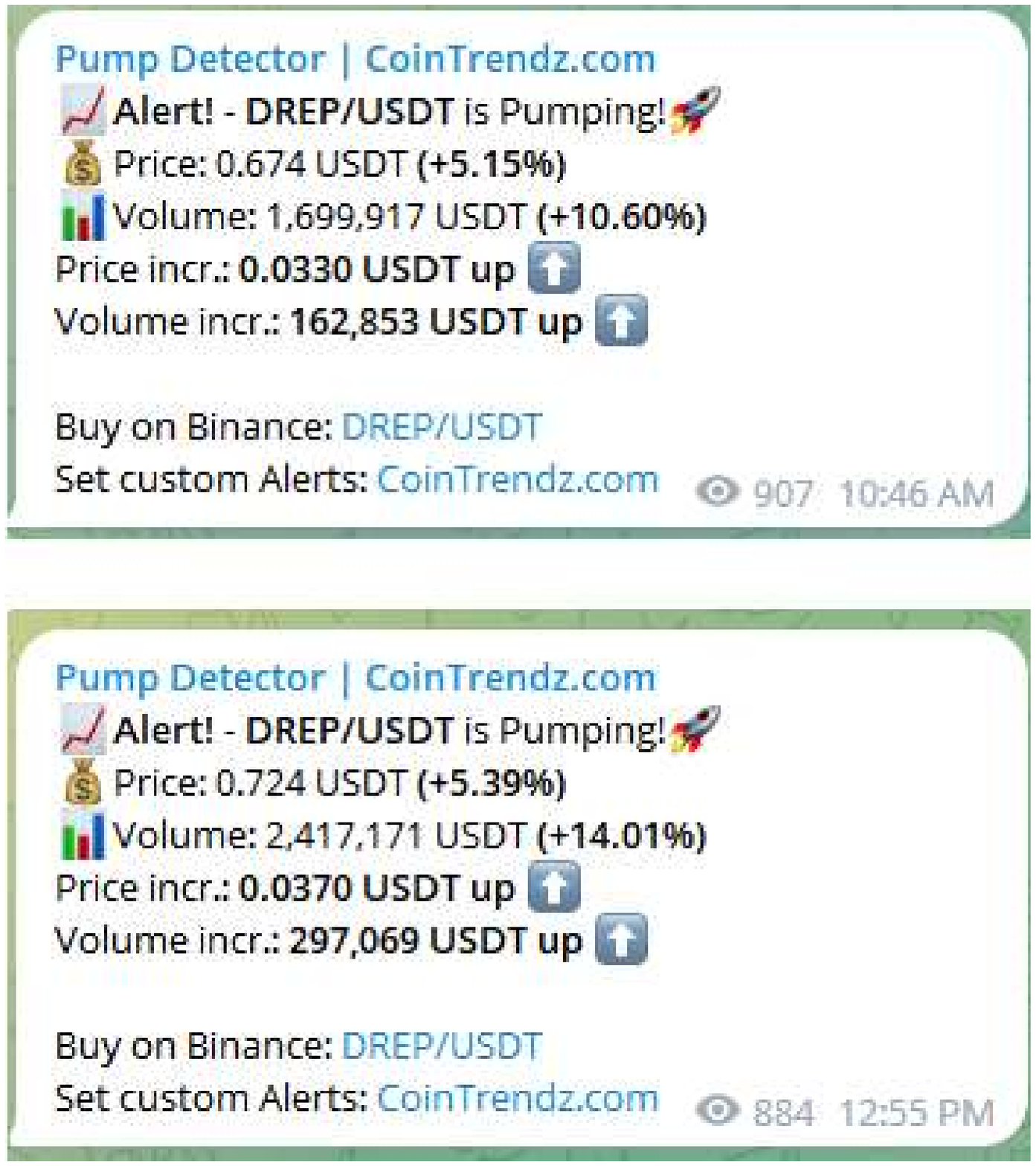}
\end{figure}

It is worth noting that other services within the Telegram infrastructure may assist in identifying cryptocurrency pump events before their occurrence, owing to price and volume fluctuations initiated by VIP members. For instance, the ``cointrendz\_pumpdetector'' channel (see Figure~\ref{f:PumpDetectorDREP}) monitors the market continuously, detecting unusual activity and pumps in multiple exchanges. This service issued two alerts related to DREP before the pump events, one five minutes before the event and another two hours and 14 minutes prior. However, it is uncertain whether the pump will indeed materialize, thus utilizing such a service may entail risks for non-VIP members. Notably, in this particular detection, the identified pair was DREP/USDT instead of DREP/BTC, as specified in the pump announcement. As USDT is a primary currency for transactions, it is possible that VIP members were either purchasing DREP using USDT obtained from fiat money (e.g., US dollars) or selling BTC for USDT and subsequently purchasing DREP.

\insertfloat{Figure~\ref{f:PumpDetectorDREP}}

\section{Data}
\label{sec:data}

\noindent
This section provides an overview of the data utilized in this study and outlines the process of constructing 
our pump-and-dump events database.

\subsection{Pump-and-Dump Groups and Events}

\noindent
To assemble a comprehensive dataset, we initially compiled a list of approximately 100 pump-and-dump groups sourced from various forums on platforms such as Reddit and Discord. Additionally, we consulted websites that curate lists of cryptocurrency Telegram channels.\footnote{For instance, refer to \url{https://coingape.com/best-telegram-crypto-channels-list/}.} Subsequently, we conducted an extensive search within the messages of these groups to identify links to other relevant groups, thereby expanding our list to encompass more than a hundred distinct groups.

The Telegram API, which is publicly accessible, facilitated downloading all messages from the groups or channels we had subscribed to or joined. Consequently, we extracted all messages along with their corresponding publication dates. To detect group-specific patterns, we implemented keyword searches within user conversations. However, it is worth noting that communication styles vary across groups. For example, some groups restrict participation solely to administrators, while others foster open discussions among multiple users. Although groups exhibit heterogeneity in their communication dynamics, certain groups adhere to well-defined patterns: (1) an announcement of a future event, (2) frequent reminders, and (3) the dissemination of a buy signal. Based on these discernible patterns, we manually compiled the events.

Our data collection focuses exclusively on events occurring on the Binance exchange using the Bitcoin (BTC) trading pair. We justify this choice by postulating that Bitcoin, as a highly speculative asset, is unlikely to be the primary target of pump-and-dump schemes. While Bitcoin may have been targeted in the past, our study period did not uncover any events involving this cryptocurrency. Furthermore, we limit our scope to the Binance exchange due to access limitations, whereby we only have historical minute-level price data from Binance rather than multiple competing exchanges. Moreover, we exclude events related to Dogecoin, Litecoin, Ethereum, and Cardano, as their widespread popularity posed challenges in collecting data from Twitter, given the limitations on the number of tweets we could collect. In total, our dataset comprises 1,160 pump-and-dump events.

\subsection{Price and Volume Data}

\noindent
We gather volume and closing price data for cryptocurrencies at a one-minute frequency from the website \url{www.cryptoarchive.com.au}. This data originates from the Binance API, ensuring that our analysis reflects the price and volume information available on the Binance trading platform.

\subsection{Twitter Data}

\noindent
Twitter is a microblogging social network enabling users to share posts with a maximum length of 280 characters. As of 2023, the platform boasts more than 450 million active users, and it was reported in 2020 that over 500 million messages are posted daily. To obtain Twitter data relevant to our study, we employ the Twitter Search API. Specifically, we search for tweets associated with each event, employing either the event's hashtag (\eg, \#BTC) or the cashtag (\eg, \$BTC) corresponding to the targeted cryptocurrency. For each event, we download tweets spanning a period from two days before the event to 12 hours after the event's scheduled time. Our focus primarily lies on tweets posted in English, and we include retweets in our dataset. Furthermore, we exclude tweets containing more than five hashtags or cashtags. This criterion allows us to filter out tweets that attempt to garner increased visibility by utilizing unrelated cashtags and hashtags. Such tweets often lack a specific association with the underlying cryptocurrency of interest, that is, the one expected to undergo a pump-and-dump event in the immediate future.\footnote{For example, consider the tweets in the following link: \url{https://twitter.com/CoinLegs/status/1381557851697582087}. These tweets reference 24 different cryptocurrencies and do not necessarily pertain to a specific cryptocurrency that will experience a pump-and-dump event in the near future.}

\section{Empirical Analysis}
\label{sec:empirical}

\noindent
This section employs an event study methodology to examine the dynamics surrounding pump-and-dump events. The primary objective is to investigate whether an abnormal level of Twitter activity preceding the event is associated with any abnormal return movement during the event, including the periods before, during, and after the pump-and-dump event. To accomplish this, we first refine our event database to focus on ``successful pump-and-dump events.'' We then define the variable used in the event study and conduct our empirical analysis to address two key questions: (1) Is there promotion of pump-and-dump events on Twitter? (2) Does Twitter promotion influence the return dynamics surrounding the pump-and-dump event?

\subsection{Qualifying the Success of Events}

\begin{figure}[H]
\caption{\textbf{Example of an Unsuccessful Pump-and-Dump Event}\\
This figure displays the discussion around an unsuccessful event.}
\hspace*{2.5cm}
\label{f:failedPD}
\centering
\includegraphics[scale=0.75]{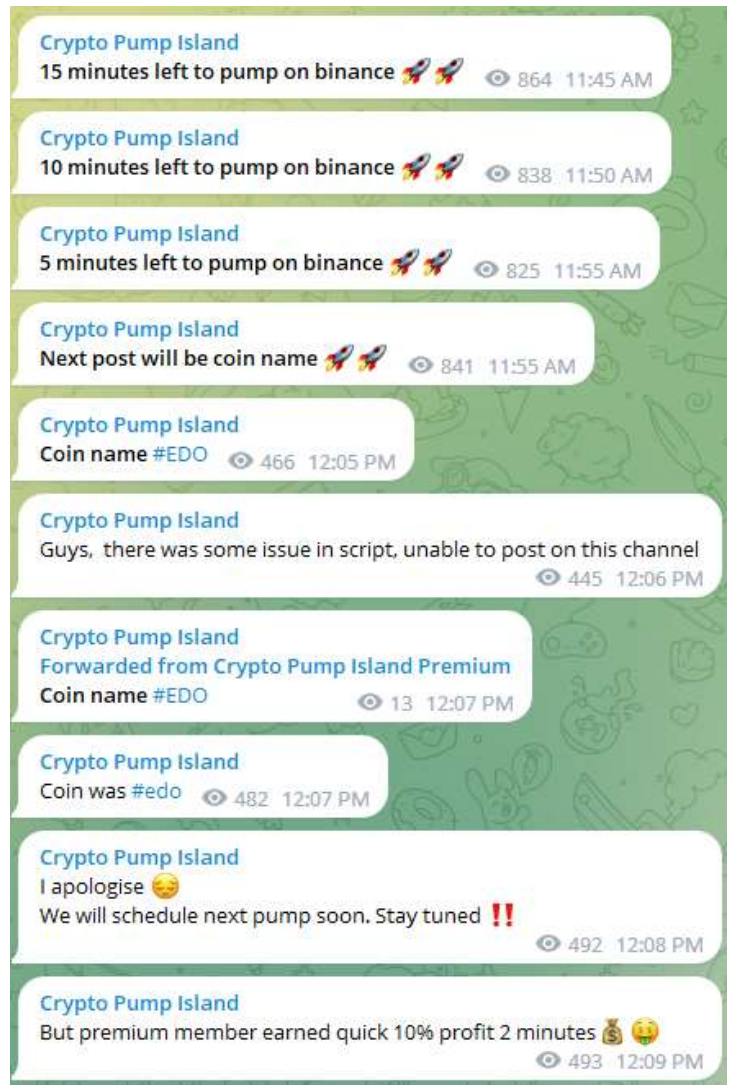}
\end{figure}

\begin{figure}[H]
\caption{\textbf{Market Reaction Around an Unsuccessful Pump-and-Dump Event}\\
This figure reports the cumulative abnormal return and the cumulative abnormal volume of ``EDO'' around its January 8, 2020, pump-and-dump events. Minute 0 is the exact announcement time reported on the ``crypto\_pump\_island'' channel, at 5:05 pm GMT.}
\label{f:failedPDreturn}
\centering
\includegraphics[scale=0.35]{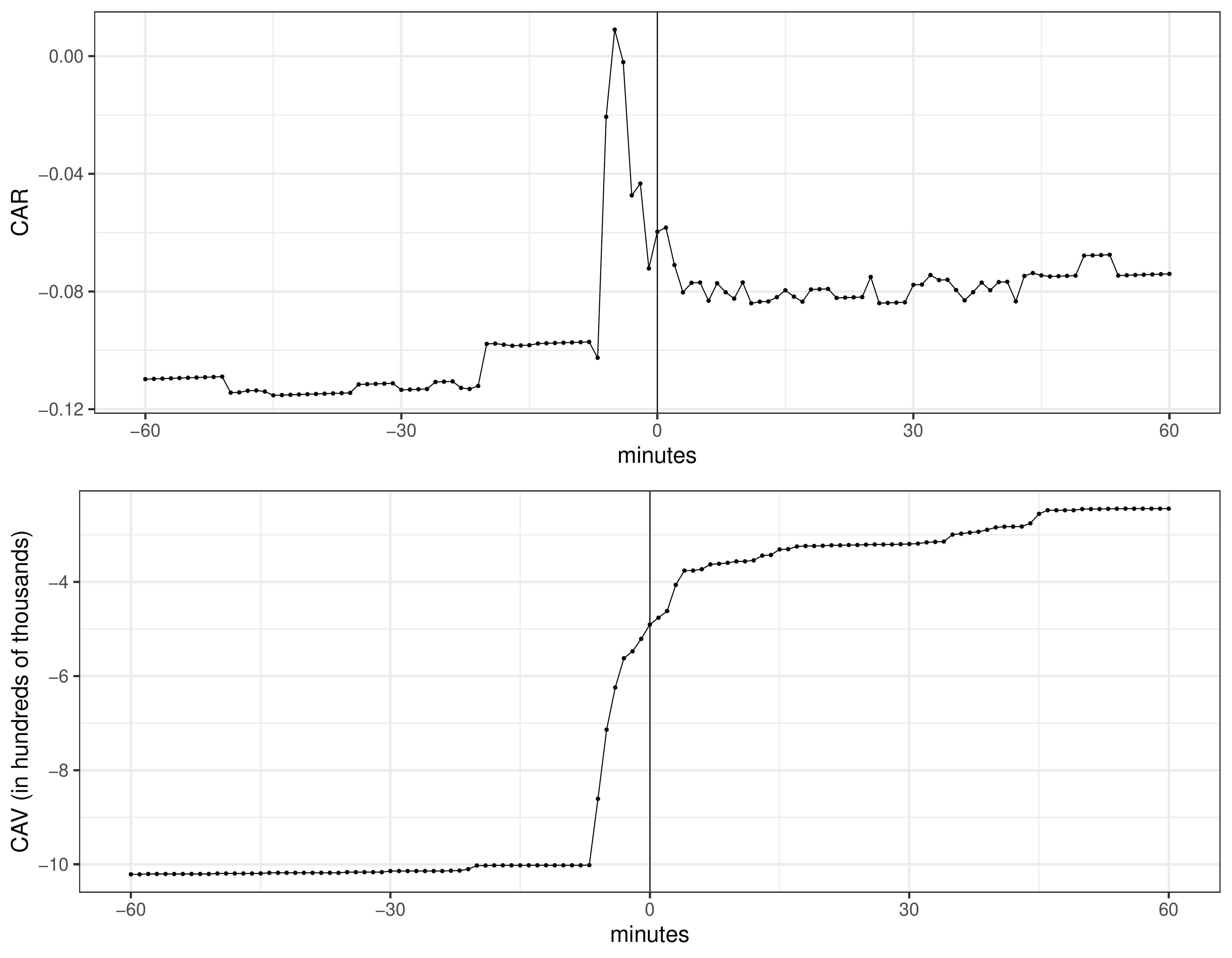}
\end{figure}

\noindent
A pump-and-dump event can be classified as either successful or unsuccessful. Several factors can contribute to a failure, such as organizational errors, insufficient participation from active members during the pump stage, or a lack of synchronization. For instance, consider the pump-and-dump event publicized by the channel \url{https://t.me/crypto_pump_island} on January 8, 2020, at 17:05 GMT, involving the cryptocurrency EDO. Figure \ref{f:failedPD} displays the chat, which reveals an error in posting the coin name. Examining the abnormal return and abnormal volume movement of the EDO/BTC pair in Figure \ref{f:failedPDreturn}, we observe that the price spike occurred five minutes before the event announcement. This observation suggests the presence of VIP users but minimal participation from non-VIP members at the time of the announcement. After the event, we note a marginal increase in volume and price, as indicated in the Crypto Pump Island chat in Figure \ref{f:failedPD}. Consequently, this event would be classified as a failure.

\insertfloat{Figure~\ref{f:failedPD} and Figure~\ref{f:failedPDreturn}}

Our strategy to identify successful events and determine the precise start time of an event involves two steps. First, we ascertain the success of an event. For each event, we collect volume data for one day before and one day after the event.\footnote{We choose a two-day span around the event as it is uncommon for a cryptocurrency to be targeted for a pump-and-dump scheme within a single day of events concerning that cryptocurrency.} Subsequently, we aggregate the minute-by-minute data into five-minute intervals, with the event minute serving as the starting point.\footnote{For example, if the event starts at 12:07, the initial five-minute chunk that sets the reference for subsequent chunks spans from 12:07 to 12:11.} If the volume within the five-minute interval encompassing the event ranks among the top three highest five-minute volumes over the two-day period (equating to 0.001\% of all observations), we classify the event as successful. In other words, if there is a significant abnormal volume during and shortly after the event announcement, indicating the participation of non-VIP members, we deem the event successful. Overall, this procedure identifies 322 successful events.

\begin{figure}[H]
\caption{\textbf{Number of Successful Pump-and-Dump Event Per Week}\\
This figure displays the number of weekly successful pump-and-dump events in our database from February 2019 to February 2022.}
\label{f:Events_per_weeks}
\centering
\includegraphics[scale=0.35]{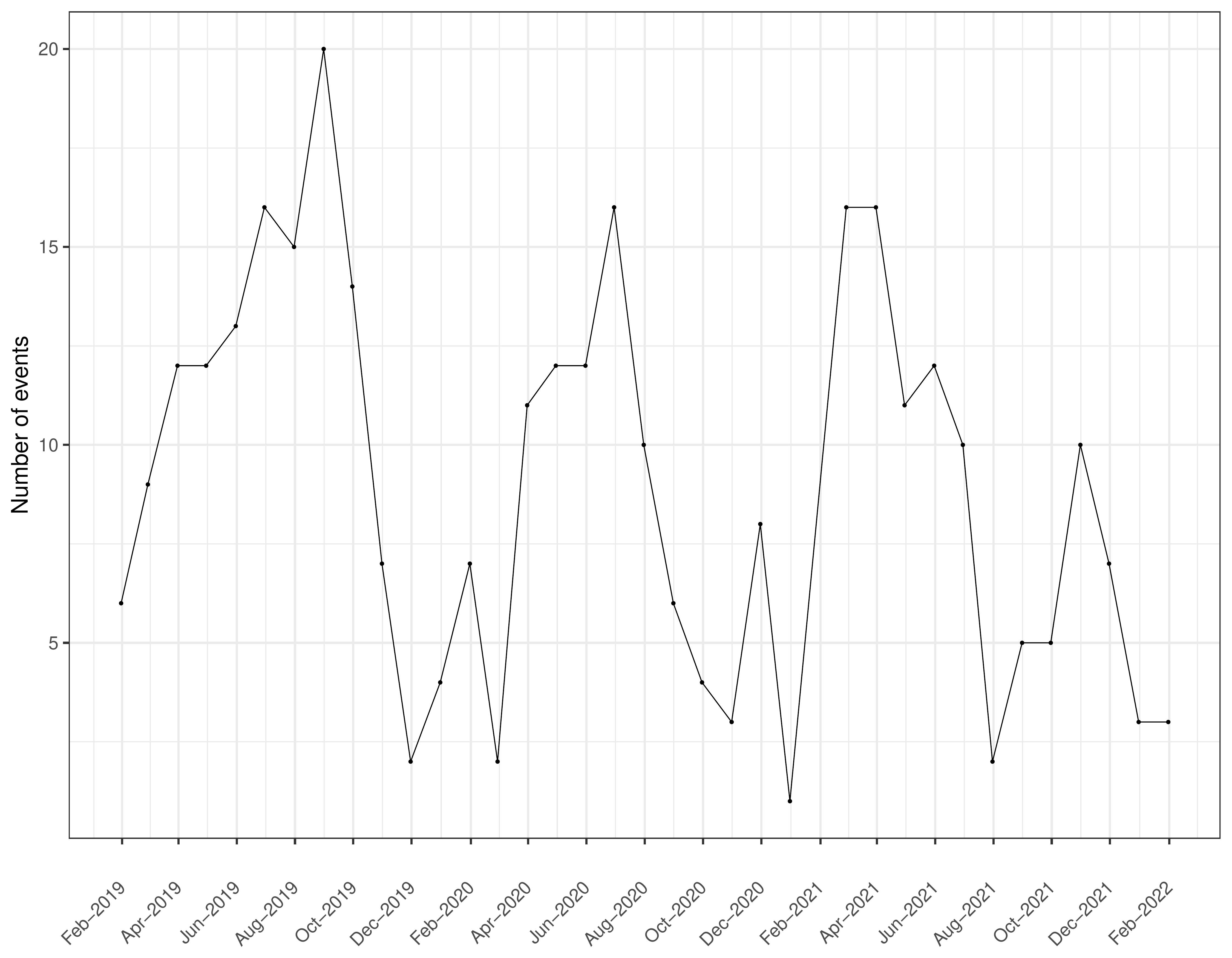}
\end{figure}

In Figure \ref{f:Events_per_weeks}, we present the number of events per week. Our dataset spans from February 2, 2019, to February 2, 2022, and exhibits considerable variability in the frequency of pump-and-dump events. Each week contains at least one event, ensuring a well-distributed dataset over time and mitigating the potential influence of specific time periods on subsequent analyses.

\insertfloat{Figure~\ref{f:Events_per_weeks}}

Table \ref{t:eventsCrypto} displays the number of events per cryptocurrency and the corresponding percentage across our dataset. No single cryptocurrency dominates significantly as a target for pump-and-dump events.

\insertfloat{Table~\ref{t:eventsCrypto}}

\begin{figure}[H]
\caption{\textbf{Number of Successful Event Per Cryptocurrencies}\\
This figure displays a histogram of the number of successful events in our database per cryptocurrency.}
\label{f:events_per_crypto}
\centering
\includegraphics[scale=0.35]{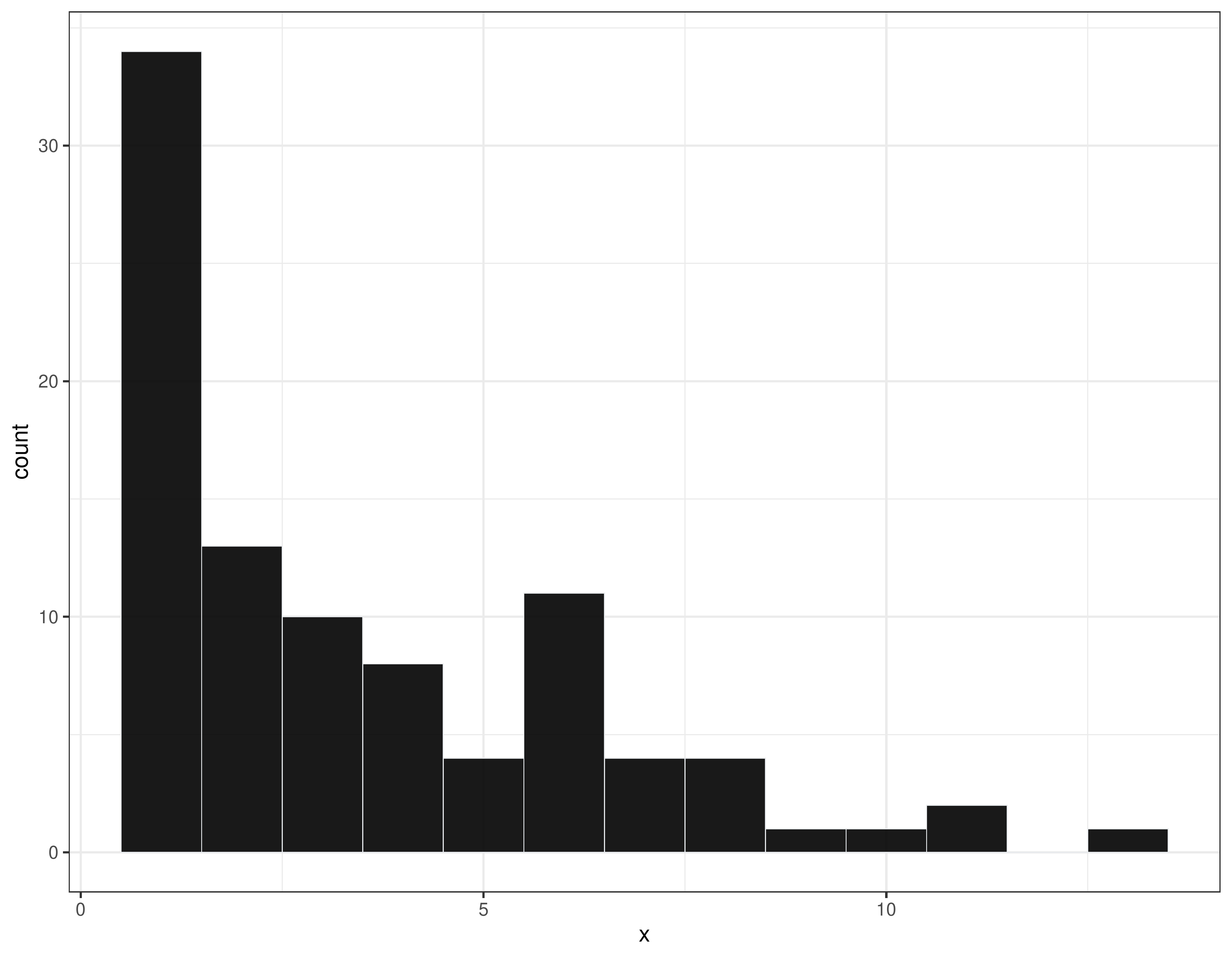}
\end{figure}

To supplement the information provided in Table \ref{t:eventsCrypto}, Figure \ref{f:events_per_crypto} presents a histogram illustrating the distribution of the number of events per cryptocurrency. While Table \ref{t:eventsCrypto} indicates that no cryptocurrency exhibits significant dominance in our database, it is evident that several of the 93 unique cryptocurrencies featured in our dataset (of which there was a successful event) have only been associated with a single event. The average number of events per cryptocurrency is 3.46, with a median of 2.

\insertfloat{Figure~\ref{f:events_per_crypto}}

Table \ref{t:tweetsStatistics} presents summary statistics regarding the number of tweets surrounding a pump-and-dump event. On average, there are 187.81 tweets associated with each event, accompanied by a substantial standard deviation of 1,119.05. The median number of tweets is 15, indicating a highly skewed distribution.

\insertfloat{Table~\ref{t:tweetsStatistics}}

\subsection{Event-Study Methodology}
\label{sec:methodology}

\noindent
This section presents the methodology employed to analyze pump-and-dump events, focusing on identifying and measuring abnormal returns, volume, and tweets. Additionally, the concept of cumulative measure is introduced to assess the cumulative impact of these factors during specific time intervals. This section closely follows the typical event study setup performed in the economic and financial literature \citep{mackinlay1997event}.

\subsubsection{Event Periods}

\noindent
We divide the analysis into five periods to capture the dynamics surrounding pump-and-dump events. First, the \textit{training period} encompasses the two days preceding the event up to 12 hours before its occurrence. This period serves as a reference for estimating ``normal'' market behavior. The \textit{pre-event period} extends from 12 hours (720 minutes) before the event to 1 minute before its initiation and is characterized by the possibility of VIPs initiating purchases of the targeted cryptocurrency. The subsequent period is the \textit{pump window}, spanning from 1 minute before the event to 1 minute after it. During this phase, non-VIP members receive the signal from the event organizers and start purchasing the cryptocurrency. The \textit{dump window} follows the pump window and lasts from 1 to 31 minutes after the event. In this interval, both VIPs and non-VIPs may commence selling their holdings. Finally, the \textit{post-dump window} extends from 32 minutes to 12 hours after the event and allows for further trading activity, including selling by late participants and potential buying by new investors. Notably, the pump's effects may not be wholly reversed during the dump window, indicating the presence of less efficient or informed investors. This occurs when less knowledgeable participants sell late after the price peak, reflecting their lack of understanding of the event dynamics.

\subsubsection{Abnormality Measures}

\noindent
The analysis of pump-and-dump events focuses on abnormal returns, abnormal volume, and abnormal tweets as market activity indicators. The computation process for abnormal returns and abnormal volume is similar. For a given event $i$ and a specific time $t$, the abnormal return $\textit{AR}_{i,t}$ is defined as the difference between the observed return $r_{i,t}$ and the expected return $E[r_{i,t}]$:
\begin{equation}\label{eq:abnormal_return}
\textit{AR}_{i,t} = r_{i,t} - E[r_{i,t}]\,,
\end{equation}
where $r_{i,t}$ represents the one-minute logarithmic return of the cryptocurrency targeted by event $i$ at time $t$. Similarly, the abnormal volume $\textit{AV}_{i,t}$ is computed as the difference between the observed volume $V_{i,t}$ and the expected volume $E[V_{i,t}]$:
\begin{equation}\label{eq:abnormal_volume}
\textit{AV}_{i,t} = V_{i,t} - E[V_{i,t}]\,,
\end{equation}
with $V_{i,t}$ denoting the one-minute volume of the targeted cryptocurrency at time $t$. The expected return $E[r_{i,t}]$ and expected volume $E[V_{i,t}]$ are estimated based on the average one-minute log return and volume, respectively, calculated using the training period data.

Abnormal tweets, $\textit{AT}_{i,t}$, are determined by comparing the number of tweets $T_{i,t}$ discussing the targeted cryptocurrency at time $t$ with the expected number of tweets $E[T_{i,t}]$. Here also, the expected value is estimated based on the average number of tweets in one minute on the training window. The abnormal tweets are then normalized by dividing the difference by the standard deviation $\sigma_{T_{i,t}}$ of the number of tweets published during the training period:
\begin{equation}\label{eq:abnormal_tweets}
\textit{AT}_{i,t} = \frac{T_{i,t} - E[T_{i,t}]}{\sigma_{T_{i,t}}}\,.
\end{equation}
The normalization of abnormal tweets accounts for the significant variability in the number of tweets published across different cryptocurrencies. It is important to note that the normalization does not qualitatively affect the results.

\subsubsection{Cumulative Abnormality}

\noindent
In addition to assessing abnormality at specific time points, we define cumulative measures to assess the cumulative impact of abnormal returns, volume, and tweets over certain time intervals. For a given event $i$, with the pump announcement time set as $t=0$, the cumulative abnormal return ($CAR_i$) between relative times $\tau_1$ and $\tau_2$ is calculated as the sum of the abnormal returns $AR_{i,\tau}$ over this interval:
\begin{equation}\label{eq:abnormal_return}
\textit{CAR}_i(\tau_1,\tau_2) = \sum_{\tau=\tau_1}^{\tau_2} \textit{AR}_{i,\tau}\,.
\end{equation}
Similarly, the cumulative abnormal volume ($\textit{CAV}_i$) and cumulative abnormal tweets ($\textit{CAT}_i$) are computed as the sums of the abnormal volume $\textit{AV}_{i,\tau}$ and abnormal tweets $\textit{AT}_{i,\tau}$, respectively, over the specified time interval:
\begin{align}\
\textit{CAV}_i(\tau_1,\tau_2) &= \sum_{\tau=\tau_1}^{\tau_2} \textit{AV}_{i,\tau}\,,\\
\textit{CAT}_i(\tau_1,\tau_2) &= \sum_{\tau=\tau_1}^{\tau_2} \textit{AT}_{i,\tau}\,.
\end{align}
These cumulative abnormality measures allow us to analyze the aggregated impact of abnormal returns, volume, and tweets over specific periods, providing insights into the overall market behavior surrounding pump-and-dump events.

\subsection{Graphical Analysis}

\begin{figure}[H]
\caption{\textbf{Graphical Analysis of the Pump-and-Dump Event Study}\\
This figure reports the average cumulative abnormal return (CAR), average cumulative abnormal volume (CAV), and average cumulative abnormal tweets (CAT) across all successful events in our dataset. Minute 0 is the identified announcement time.}
\label{f:eventStudy}
\centering
\includegraphics[scale=0.35]{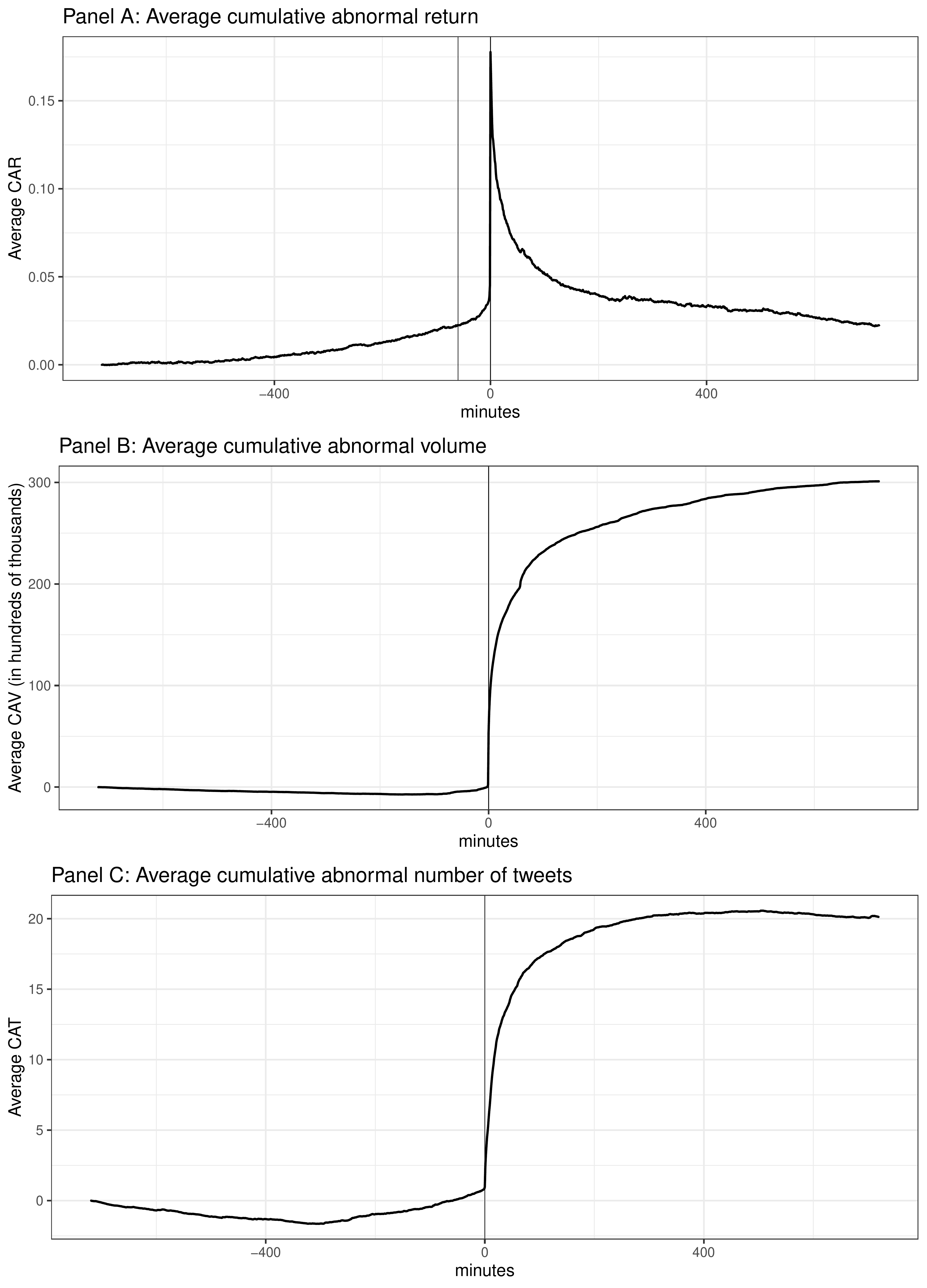}
\end{figure}

\noindent
In this section, we present a graphical analysis to examine the evolution of cumulative abnormal return, volume, and tweets over a range of relative time intervals. Specifically, we start from $\tau_1 = -720$ (12 hours before) and increment by one minute to $\tau_2$, covering the period from $\tau_2 = -720$ to $\tau_2 = 720$. The graphical representations are displayed in Figure \ref{f:eventStudy}, with Panel A, B, and C illustrating the average cumulative abnormal return, volume, and tweets, respectively.

Our analysis reveals several noteworthy patterns. First, we observe a positive and increasing trend in abnormal returns during the pre-event window (from 12 hours before the event to 1 minute before). This upward trend becomes more pronounced towards the end of the window, approximately 1 hour before the event. As for abnormal volume, we observe an overall decreasing trend, with a shift towards positive values also evident towards the latter part of the pre-event window, around 2 hours before the event. In terms of tweets, a significant negative trend is observed approximately 5 hours before the event, followed by a shift towards a positive trend. These observations suggest that successful pump-and-dump events tend to be initiated on cryptocurrencies that exhibit positive trends, accompanied by abnormal trading volume and promotion on Twitter.

We observe a substantial spike in abnormal return, volume, and tweets during the pump window. This surge reflects the involvement of both VIP and non-VIP users and may indicate further promotion or increased awareness of the event on Twitter.

In the dump window, we observe a rapid and significant decrease in abnormal returns, indicating the dumping of the targeted cryptocurrency. The positive trend in abnormal volume further supports this observation. Interestingly, abnormal positive activity on Twitter persists during this window, which could be related to the dump itself or an attempt to attract additional buyers.

Moving to the post-dump window, we still observe a negative trend in abnormal returns, although to a lesser extent than the dump window. However, a positive trend in abnormal volume accompanies this decline. These findings suggest the presence of participants who continue to sell their cryptocurrency during this stage, which can be considered ``late'' participation given the nature of the event. Regarding tweets, we observe a positive trend until approximately 5 hours after the event, after which the trend becomes flat, indicating the absence of further abnormal activity on Twitter.

\insertfloat{Figure~\ref{f:eventStudy}}

Based on these results, we can conclude that Twitter serves as a platform for promoting pump-and-dump events. However, it is important to note that this analysis does not conclusively establish whether such promotion successfully attracts additional participants to engage in the pump-and-dump event. This aspect will be further investigated in our subsequent analysis.

\subsection{Regression Analysis}

\noindent
We posit that promoting a cryptocurrency targeted by a pump-and-dump event on Twitter can attract investors, primarily due to increased attention directed towards that specific cryptocurrency. In the presence of promotion, we anticipate a contemporaneous relationship between the abnormal number of tweets before the event and the abnormal return leading up to the event.

Some investors may not be privy to the exact timing of the pump and subsequently find themselves delayed in selling their tokens during the dump phase. These investors would only become aware of the dump when examining their portfolios. Consequently, even if they purchased the cryptocurrency before the pump, their ability to profitably sell their tokens may be compromised due to their late entry into the dump phase. In this scenario, we expect to observe a relationship between the abnormal number of tweets before the event and the post-dump abnormal return, with the post-dump period commencing 30 minutes after the pump window.

Alternatively, these investors could possess full awareness of the nature of the event, providing them with an advantage over other participants (non-VIPs). In such cases, we anticipate a relationship between the abnormal number of tweets before the event and the abnormal return during the dump phase, which encompasses the period up to 30 minutes after the pump.

We employ regression analysis to ascertain whether Twitter promotion effectively attracts additional participants to engage in the pump-and-dump event. Our analysis starts by investigating the existence of a contemporaneous relationship between the abnormal number of tweets preceding the event and the cumulative abnormal return leading up to the event. Specifically, we focus on the last 30 minutes of the pre-event window. To conduct this analysis, we estimate the following regression model:
\begin{equation}\label{eq:abnormal_return}
\textit{CAR}_i(-31,-2) = \alpha + \beta \textit{CAT}_i(-31,-2) + \varepsilon_i\,.
\end{equation}
All variables in our analysis, except for the one about the cumulative abnormal return, are standardized to have a mean of 0 and a standard deviation of 1. The regression results are presented in the first column of Table~\ref{t:regression}.\footnote{It is worth noting that incorporating a cryptocurrency fixed effect did not significantly alter the qualitative outcomes of our analysis. However, due to the limited frequency of most cryptocurrencies in our database, including a cryptocurrency fixed effect would impose substantial constraints on our analysis.}

Consistent with the notion that promotion on Twitter engenders heightened attention towards the targeted cryptocurrency, subsequently generating increased buying pressure, we observe a positive and statistically significant relationship between the abnormal number of tweets and abnormal returns. Specifically, during the pre-event window, a one standard deviation increase in the cumulative abnormal number of tweets corresponds to a contemporaneous increase of 0.325\% in cumulative abnormal returns.

\insertfloat{Table~\ref{t:regression}}

Subsequently, we investigate the influence of the cumulative abnormal number of tweets before the event on the abnormal returns during the dump phase. As previously mentioned, if investors informed via Twitter about the pump-and-dump event exhibit awareness of its nature, we anticipate a negative relationship between the abnormal number of tweets before the event and the abnormal return during the dump phase. Conversely, in the alternate scenario where investors are unaware, we expect the relationship to be statistically insignificant. To explore this, we estimate the following regressions:
\begin{equation}\label{eq:abnormal_return}
\textit{CAR}_i(2,31) = \alpha + \beta \textit{CAT}_i(-31,-2) + \varepsilon_i\,,
\end{equation}
and an additional regression where we incorporate control variables related to the 
pump-and-dump event environment:
\begin{multline}\label{eq:abnormal_return}
\textit{CAR}_i(2,31) = \alpha 
+ \beta \textit{CAT}_i(-31,-2) 
+ \lambda_1 \textit{CAT}_i(-1,1) \\
+ \lambda_2 \textit{CAR}_i(-31,-2) 
+  \lambda_3 \textit{CAR}_i(-1,1) \\
+ \lambda_4 \textit{CAV}_i(-31,-2) 
+ \lambda_5 \textit{CAV}_i(-1,1) 
+ \varepsilon_i\,.
\end{multline}

Columns 2 and 3 of Table~\ref{t:regression} present the results of these two regressions. Our findings reveal a significant negative relationship between the cumulative abnormal return at and preceding the event and its equivalent during the dump window. This indicates that a substantial pump event corresponds to a sizable dump event. However, we do not observe a significant relationship between the abnormal number of tweets before the event and the cumulative abnormal return during the dump window. This outcome suggests that investors informed via Twitter are not selling their investments during the dump period.

Finally, we examine whether the cumulative abnormal number of tweets before the event impacts abnormal returns during the post-dump phase. Considering the absence of a relationship between these variables during the dump period, we now expect that investors who receive information from Twitter will sell their cryptocurrencies after the dump period. Hence, we anticipate a negative relationship between the cumulative abnormal number of tweets before the event and the cumulative abnormal return during the post-dump phase. To explore this, we estimate the following regressions:
\begin{equation}\label{eq:abnormal_return}
\textit{CAR}_i(31,720) = \alpha + \beta  \textit{CAT}_i(-31,-2) + \varepsilon_i,,
\end{equation}
and, similar to the previous analysis, a regression with additional control variables:
\begin{multline}\label{eq:abnormal_return}
\textit{CAR}_i(31,720) 
= \alpha + \beta  \textit{CAT}_i(-31,-2) 
+ \lambda_1  \textit{CAT}_i(-1,1) 
+ \lambda_2  \textit{CAT}_i(2,31) \\ 
+ \lambda_3  \textit{CAR}_i(-31,-2) 
+ \lambda_4  \textit{CAR}_i(-1,1) 
+ \lambda_5  \textit{CAR}_i(2,31) \\ 
+ \lambda_6  \textit{CAV}_i(-31,-2) 
+ \lambda_7  \textit{CAV}_i(-1,1) 
+ \lambda_8  \textit{CAV}_i(2,31) + \varepsilon_i\,.
\end{multline}

Columns 4 and 5 of Table~\ref{t:regression} present the results of these regressions. Notably, we observe a predominantly negative and statistically significant relationship between the post-dump abnormal return and the cumulative abnormal number of tweets before the event. This indicates that the price increase observed before the event, which is associated with abnormal tweets, only reverses at least 30 minutes after the dump phase. This finding suggests that investors who receive information from Twitter regarding a pump-and-dump event tend to be delayed in selling their cryptocurrencies compared to investors who receive information through other channels related to the event's announcement. Moreover, after controlling for other variables, a one standard deviation increase in the number of abnormal tweets before the event leads to a negative cumulative abnormal return of -2.202\%. This reversal is substantially larger than the contemporaneous effect, implying that investors informed via Twitter who are late in selling their cryptocurrencies during a pump-and-dump event tend to experience, on average, a loss on their investment.

Overall, our regression analysis provides evidence that the cumulative abnormal number of tweets before the event influences the abnormal returns before, during, and after the pump-and-dump event. These results support the notion that Twitter promotion attracts participants to the event and influences their trading behavior.

\section{Conclusion} 
\label{sec:conclusion}

\noindent
Our study provides compelling evidence that Twitter promotion is pivotal in attracting participants to pump-and-dump events in the cryptocurrency market. Analyzing abnormal returns, trading volume, and tweets offers valuable insights into the temporal dynamics of these events. Our regression analysis further confirms the influence of Twitter activity on abnormal returns before, during, and after the event, highlighting the significance of social media in shaping participant behavior. These findings contribute to a deeper understanding of the mechanisms underlying pump-and-dump events and emphasize the need for increased regulatory measures to mitigate the risks of market manipulation facilitated through social media platforms.

\newpage
\singlespacing
\bibliography{ref}

\newpage
\begin{table}[H]
\centering
\caption{\textbf{Number of Events Per Cryptocurrency}\\
This table reports the 20 cryptocurrencies with the largest number of pump-and-dump events. It shows the number of events and the percentage of occurrence among all events.}
\label{t:eventsCrypto}
\begin{tabular}{lcc}
\toprule
Cryptocurrency & Events & \% \\ 
\midrule
BRD &  13 & 4.04 \\ 
NEBL &  11 & 3.42 \\ 
NXS &  11 & 3.42 \\ 
OAX &  10 & 3.11 \\ 
RDN &   9 & 2.80 \\ 
GVT &   8 & 2.48 \\ 
NAV &   8 & 2.48 \\ 
VIA &   8 & 2.48 \\
VIB &   8 & 2.48 \\ 
APPC &   7 & 2.17 \\ 
BNT &   7 & 2.17 \\ 
QSP &   7 & 2.17 \\ 
SNM &   7 & 2.17 \\ 
BLZ &   6 & 1.86 \\ 
CDT &   6 & 1.86 \\ 
CTXC &   6 & 1.86 \\ 
DLT &   6 & 1.86 \\ 
EDO &   6 & 1.86 \\ 
GRS &   6 & 1.86 \\ 
HC &   6 & 1.86 \\ 
\bottomrule
\end{tabular}
\end{table}

\newpage
\begin{table}[H]
\centering
\caption{\textbf{Number of Tweets Around Events}\\
This table shows the statistics of the number of tweets around pump-and-dump events (from 2 days before to 12 hours after the announcement time of each event). The statistics include the mean, standard deviation, minimum, 25th percentile, median, 75th percentile, and maximum number of tweets.}
\label{t:tweetsStatistics}
\begin{tabular}{lr}
\toprule
Statistics & Tweets \\
\hline
Mean & 187.81 \\
Std & 1,119.05 \\
Min & 0 \\
25th & 5 \\
50th & 15 \\
75th & 53 \\
Max & 20,439 \\
\bottomrule
\end{tabular}
\end{table}

\newpage
\begin{table}[H]
\centering
\caption{\textbf{Regression Results}\\
This table reports the various regression results. Figures in parentheses denote the standard deviation of the estimated differences. $^{***}$, $^{**}$, $^{*}$ indicate that the coefficient is significantly different from zero at the 1\%, 5\%, and 10\% levels.}
\label{t:regression}
\scalebox{0.95}{
\begin{tabular}{lccccc}
\toprule
& \textit{CAR}(-31,-2) 
& \textit{CAR}(2,31) 
& \textit{CAR}(2,31) 
& \textit{CAR}(32,720) 
& \textit{CAR}(32,720) \\ 
\midrule
Intercept & -0.008 & -9.752$^{***}$ & 0.548 & -5.796$^{***}$ & -3.629$^{***}$ \\ 
& (0.042) & (0.854) & (0.702) & (0.533) & (0.723) \\ 
\midrule
\textit{CAT}(-31,-2) & 0.325$^{***}$ & -0.47 & 0.257 & -2.202$^{***}$ & -1.256$^{**}$ \\ 
& (0.049) & (0.991) & (0.646) & (0.618) & (0.62) \\ 
\textit{CAT}(-1,1) &  &  & -0.127 &  & 0.836 \\ 
&  &  & (0.681) &  & (0.641) \\ 
\textit{CAT}(2,31) &  &  &  &  & -3.077$^{*}$ \\ 
&  &  &  &  & (1.684) \\ 
\midrule
\textit{CAR}(-31,-2) &  &  & -2.979$^{***}$ &  & -1.718$^{**}$ \\ 
&  &  & (0.701) &  & (0.685) \\ 
\textit{CAR}(-1,1) &  &  & -0.732$^{***}$ &  & -0.279$^{***}$ \\ 
&  &  & (0.034) &  & (0.058) \\ 
\textit{CAR}(2,31) &  &  &  &  & -0.167$^{***}$ \\ 
&  &  &  &  & (0.058) \\ 
\midrule
\textit{CAV}(-31,-2) &  &  & -0.654 &  & -1.816$^{**}$ \\ 
&  &  & (0.748) &  & (0.722) \\ 
\textit{CAV}(-1,1) &  &  & -0.387 &  & 1.196 \\ 
&  &  & (0.629) &  & (1.18) \\ 
\textit{CAV}(2,31) &  &  &  &  & -1.517 \\ 
&  &  &  &  & (1.181) \\
\midrule
Adj. R$^2$ (\%) & 11.8 & -0.24 & 63.78 & 3.52 & 20.98 \\ 
\bottomrule
\end{tabular}}
\end{table}

\newpage

\newpage

\newpage

\newpage

\newpage

\newpage

\newpage

\newpage

\newpage

\newpage

\end{document}